\documentclass[11pt,titlepage]{article}

\usepackage{setspace} 
\usepackage{amsfonts}          
\usepackage{amssymb}           
\usepackage{amsmath}           
\usepackage{color}
\usepackage{epsfig}
\usepackage{psfrag}

\usepackage[numbers,sort&compress]{natbib}
\newcommand{\f}[2]{\frac{#1}{#2}}
\newcommand{\dd}{\partial}
\newcommand{\de}{{\rm \, d}}

\renewcommand{\vec}[1]{\mbox{\boldmath $ #1$}}

\allowdisplaybreaks

\newcount\HH
\newcount\MM
\MM=\time
\HH=\time
\divide\HH by 60
\divide\MM by 60
\multiply\MM by 60
\MM=-\MM
\advance\MM by \time
\def\mytime{\number\HH:\ifnum\MM<10{}0\fi\number\MM}

\newcommand\etal{\mbox{\textit{et al.}}}

\newcommand\eg{e.g.}
\newcommand\ie{i.e.}

\newcommand{\Block}{\mathcal{B}}
\newcommand{\CM}{C_M}
\newcommand{\Curv}{\mathcal{K}}
\newcommand{\curv}{\kappa}
\newcommand{\D}{\partial}

\newcommand{\dbar}{\Qstat{d}}
\newcommand{\emb}[1]{{#1}}
\newcommand{\E}{V}              
\newcommand{\Eh}{\E_h}
\newcommand{\Em}{\E_m}
\newcommand{\ENa}{\E_{Na}}
\newcommand{\Ealpha}{\E_{\alpha}}
\newcommand{\Eomega}{\E_{\omega}}
\newcommand{\Ecr}{\E_{*}}
\newcommand{\g}{g}              
\newcommand{\gNa}{g_{Na}}
\newcommand{\Heav}{\theta}
\renewcommand{\hbar}{\Qstat{h}}
\newcommand{\INa}{I_{\mathrm{Na}}}
\newcommand{\INabar}{\overline{\INa}}
\newcommand{\jmin}{j_{\min}}
\newcommand{\jcr}{j_{*}}
\newcommand{\Ki}{\mathrm{K}_i}
\newcommand{\mbar}{\Qstat{m}}
\newcommand{\mm}{\mathrm{mm}}
\newcommand{\ms}{\mathrm{ms}}
\newcommand{\Nai}{\mathrm{Na}_i}
\newcommand{\oabar}{\Qstat{\oa}}
\newcommand{\oa}{o_a}
\newcommand{\Qstat}[1]{\overline{#1}}
\newcommand{\SIsmall}{{\Sigma_{I}^{\prime}}}
\newcommand{\uabar}{\Qstat{\ua}}
\newcommand{\ua}{u_a}
\newcommand{\wbar}{\Qstat{w}}

\newcommand{\mx}[1]{\mathbf{#1}}
\newcommand{\U}{\mx{U}}
\newcommand{\F}{\mx{F}}

\newcommand{\eq}[1]{Eq.~\ref{#1}}
\newcommand{\eqs}[1]{Eqs.~\ref{#1}}
\def\eqtwo(#1,#2){Eqs.~\ref{#1},\ref{#2}}
\newcommand{\Eq}[1]{Equation~\ref{#1}}
\newcommand{\Eqs}[1]{Equations~\ref{#1}}
\newcommand{\fig}[1]{Fig.~\ref{#1}}
\newcommand{\Fig}[1]{Figure~\ref{#1}}

\newcommand{\myfigure}[3]{} 

\newcommand{\reva}[2][0]{#2}

\newcommand{\oth}[1]{#1}

\title{Conditions for propagation and block \\ of excitation in
an asymptotic model of  atrial tissue}

\author{Radostin~D.~Simitev\\
  Department of Mathematical Sciences,\\
  University of Liverpool, UK \\
  \and Vadim~N.~Biktashev\thanks{
           Corresponding author.  Address: 
           Department of Mathematical Sciences,
           University of Liverpool,
           Liverpool, L69 7ZL, UK,
           Tel.:~+44-151-7944004, Fax:~+44-151-7944061} \\
  Department of Mathematical Sciences,\\
  University of Liverpool, UK}

\date{Submitted: August 14, 2005 \\ 
Revised: \today{} at \mytime} 

\begin{document}

\maketitle

\abstract{
Detailed ionic models of cardiac cells are difficult for numerical
simulations because they consist of a large number of equations
and contain small parameters.
The presence of small parameters, however, may be used for asymptotic
reduction of the models.
Earlier results have shown that the asymptotics of 
cardiac equations are non-standard. Here we apply such a novel
asymptotic method to an ionic model of human atrial 
tissue  in order to obtain  a reduced but accurate model for the
description of excitation fronts.
Numerical simulations of spiral waves in atrial tissue show
that wave fronts of propagating action potentials break-up and
self-terminate. Our  model, 
in particular, yields a simple analytical criterion of propagation block,
which is similar in purpose but completely
different in nature to the `Maxwell rule' in the FitzHugh-Nagumo type models.
Our new criterion agrees with
direct numerical simulations of break-up of
re-entrant waves. 
}

\clearpage

\section{Introduction}
\label{sec:1}

Refractoriness is a fundamental characteristic of biological excitable
media, including cardiac tissues. The boundary between absolute and
relative refractoriness can be defined as the boundary between the
ability and the inability of the medium to conduct excitation waves
\cite{Krinsky-1966}.
Transient conduction block is thought to be a key
event in the initiation of re-entrant arrhythmias and in the development and
the self-perpetuation of atrial and ventricular fibrillation
\cite{Moe-1962,Weiss-etal-2000,Panfilov-Pertsov-2001,Kleber-Rudy-2004}. 
So it is important to understand well 
the immediate causes and conditions of propagation blocks
and sudden break-ups in such nonstationary regimes.
The aim of the present work is to improve this understanding
via analysis of a 
mathematical model of human atrial tissue \cite{CRN98}.

\citet{Kohl-etal-2000} distinguish two types of single-cell  cardiac
models: `membrane potential models'  and `ionic current 
models'. The membrane potential models attempt to represent cellular
electrical activity by describing, with a minimal number of equations,
the spatio-temporal course of changes in membrane potential.
Their equations are constructed using a
dynamical-systems arguments to caricature various properties and
processes of cardiac function. Examples of this type of models start
with the mathematical description of heartbeat as a relaxation
oscillator by \citet{vanderPol-vanderMark-1928} and continue to play an important
role in describing biophysical behaviour \cite{Holden-Panfilov-1997}
with the the most successful one arguably being the FitzHugh-Nagumo
equations \cite{FitzHugh-1961,Nagumo-etal-1962}, 
\begin{align}
\label{FHN}
& \dd_{T} \E = D\, \dd_{X}^2 \E + \epsilon_\E\, (\E-\E^3/3 - \g),\nonumber\\
& \dd_{T} \g = \epsilon_{\g} \, (\E+\beta -\gamma \g), 
\end{align}
where $\E$ and $\g$ are dynamical variables corresponding to the action
potential and the cardiac current gating variables, $\epsilon_{\E}$,
$\epsilon_{\g}$, $\gamma$, and $\beta$ are parameters and $D$ is a
diffusion constant. 
Further examples of such models can be found in 
\cite{%
  Pertsov-Panfilov-1981,%
  Barkley-1991,%
  Winfree-1991,%
  Aliev-Panfilov-1996%
}, among others.
An attractive feature of this approach is that, along with a reasonable
description of excitability, threshold, plateau and  refractoriness, 
it focusses on generic equations which can often be treated
analytically and their dynamical properties can be extended and
applied to very different physical, chemical or biological  problems
of similar mathematical structure. The main drawback of these models,
however, is their lack  of an explicit correspondence between model
components and constituent parts of the biological system,
\eg{} ion channels and transporter proteins.
The second type of models, the ionic current models,
attempt to model action potential (AP) 
behaviour on the basis of ion fluxes in as much detail as possible in
order to fit experimental data and predict behaviour under previously
untested conditions. 
A major breakthrough in this direction of cell
modelling was the work of \citet{Hodgkin-Huxley-1952} representing the first
complete quantitative description of the giant squid axon.
The ionic concept was applied to cardiac cells by \citet{Noble-1960,Noble-1962} 
and there are now ionic models of sinoatrial node pacemaker cells
\eg{}~\cite{Demir}, atrial myocytes \eg{}~\cite{Nygren}, Purkinje
fibres \eg{}~\cite{Varghese}, ventricular myocytes
\eg{}~\cite{BeelerReuter,LuoRudy} and cardiac
connective tissue cells \eg{}~\cite{Kohl-Noble-1996}. This
is only an incomplete list and the collection of
available models continues to expand. 
The ionic models have been
successfully applied to study  various conditions of metabolic activity
and excitation-contraction coupling, feedback mechanisms, response to
drugs, etc. For recent reviews of detailed 
ionic models, their computational aspects and applications we refer to 
the reviews of \citet{Kohl-etal-2000} and \citet{Clayton}.
However, these models are very complicated 
and have to be studied mostly numerically. 
Their numerical study is aggravated by stiffness of the equations, \ie{}
broad range of characteristic time scales  of dynamic variables 
caused by numerous small parameters of the models.

  An attractive compromise is exemplified by the model of
  \citet{Fenton-Karma-1998}, which combines the simplicity of only three
  differential equations with realistic description of (crudely) the AP shape and
  (rather nicely) the dependence of the AP duration and front propagation speed on
  the diastolic interval, i.e. `restitution curves'.  Unlike the
  earlier two-component model by \citet{Aliev-Panfilov-1996} it has a
  structure similar to that of true ionic models, and its parameters
  have been fitted to mimick properties of selected four detailed
  ventricular myocyte models. It is simpler than later proposed
  models of the same "intermediate" kind such as~\cite{Bernus-etal-2002}.
  However, this deservedly popular model has not been in any
  way "derived" from any detailed model, so it 
  is only reliable within the phenomenology on which it has been
  validated, i.e. normal or premature APs, but not propagation blocks.

The problem of conditions for propagation has an elegant
solution for the FitzHugh-Nagumo system \eqs{FHN} and its generalizations,
within an asymptotic theory exploiting the difference of 
time scales of different variables, such as
$\epsilon_{\g}\ll\epsilon_{\E}$ in case of \eqs{FHN} 
\cite{Tyson-Keener-1988}. The answer is formulated in terms of the
instantaneous values of the slow variables 
($\g$ in \eqs{FHN}), and claims that excitation will propagate
if the definite integral of the kinetic term in the right hand side
of the equation for the fast variable ($\E$ in \eqs{FHN} ),
between the lower and the upper quasi-stationary states, is positive
\cite[see eq. 4.5]{Fife-1976}. This is similar to Maxwell's 
`equal areas' rule in the theory of phase transitions \cite[see
section 9.3]{Haken-1978}.  
In case of \eqs{FHN} , this rule boils down to an inequality 
for the slow variable $\g$: excitation front will propagate
if the value of $\g$ at it is less than a certain $\g_*$.
However, FitzHugh-Nagumo type models completely misrepresent
the idiosyncratic `front dissipation' scenario by which
propagation block happens in the ionic current models
\cite{Biktashev-2002}. The reason is that small parameters
in such models appear in essentially different ways from the one
assumed by the standard asymptotic theory
\cite{Suckley-Biktashev-2003b,Biktashev-Suckley-2004}.
So, this elegant `Maxwell rule' solution
is not applicable to any realistic models.

We have developed an alternative asymptotic approach based on special
mathematical properties of the detailed ionic models, not captured by
the standard theory \cite{Arnold}.  This approach demonstrated
excellent quantitative accuracy for APs in isolated Noble-1962 model
cells \cite{Biktashev-Suckley-2004}, and correctly, on a qualitative
level, described the front dissipation mechanism of break-up of
re-entrant waves in \citet{CRN98} model of human atrial tissue,
although quantitative correspondence with the full model was poor
\cite{Biktasheva-etal-2005}.
In this paper we suggest, for the first time, a refined simplified
asymptotic model of a cardiac excitation front, which provides
numerically accurate prediction of the front propagation velocity
(within 16\,\%) and its profile (within $0.7$\,mV).  It also gives an 
analytical condition for propagation block in a re-entrant wave,
expressed as a simple inequality involving the slow inactivation gate
$j$ of the fast sodium current. The condition is in excellent agreement with
results of direct numerical simulations of the \citet{CRN98} full
ionic model of 21 partial differential equations.

The paper is organised as follows. In \S \ref{sec:2} we introduce the
simplified model equations and discuss their properties.  Analytical
solutions are presented in \S \ref{sec:3} for a piecewise linear
`caricature' version of our simplified model. Accurate numerical
results and a two-dimensional test are presented in \S \ref{sec:4}.  
The paper concludes with a discussion of results and questions open
for future studies in \S \ref{sec:5}.   

\section{Mathematical formulation of the model equations} 
\label{sec:2}

\subsection{Asymptotic reduction}
\label{ssec:2.1}

In this section we briefly summarise the asymptotic arguments of
\cite{Biktasheva-etal-2005} relevant to our present purposes. We re-write
\citet{CRN98} model in the following one-parameter form:
\begin{align}
\hspace*{-6mm}
&\D_{T}{\E} = D \left(\D_{X}^2 +\Curv \D_{X} \right){\E} - \f{\left(\epsilon^{-1}\INa(\E,m,h,j) + \SIsmall(\E,\dots)
 \right)}{\CM},                                                     \nonumber\\
&\D_{T}{m} = \frac{\big(\emb{\mbar}(\E;\epsilon)-m\big)}{\epsilon \, \tau_{m}(\E)}, \quad
         \emb{\mbar}(\E;0)=M(\E)\,\Heav(\E-\Em),          \nonumber\\
&\D_{T}{h} = \frac{\big(\emb{\hbar}(\E;\epsilon)-h\big)}{\epsilon \, \tau_{h}(\E)}, \quad
         \emb{\hbar}(\E;0)=H(\E)\,\Heav(\Eh-\E),          \nonumber\\
&\D_{T}{\ua} = \frac{\big(\uabar(\E)-\ua\big)}{\epsilon \, \tau_{\ua}(\E)}, \nonumber\\
&\D_{T}{w} = \frac{\big(\wbar(\E)-w\big)}{\epsilon \, \tau_{w}(\E)}, \nonumber\\
&\D_{T}{\oa} = \frac{\big(\oabar(\E)-\oa\big)}{\epsilon \, \tau_{\oa}(\E)},      \nonumber\\
&\D_{T}{d} = \frac{\big(\dbar(\E)-d\big)}{\epsilon \, \tau_{d}(\E)},             \nonumber\\
&\D_{T}{\U} = \F(\E,\dots)                     \label{e:CRN}
\end{align}
where $D$ is the voltage diffusion constant,
$\epsilon$ is a small parameter used for the asymptotics,
    $\Curv$ is the curvature of the propagating front,
$\Heav()$ is the Heaviside function, 
$\SIsmall()$ is the sum of all currents except the fast sodium current $\INa$,
the dynamic variables $\E$, $m$, $h$, $\ua$, $\oa$ and $d$
are defined in \cite{CRN98}, 
$\U=(j,o_i,\dots,\Nai,\Ki,\dots)^T$ is the vector of all other, slower variables, 
and $\F$ is the vector of the corresponding right-hand sides. 
The rationale for this parameterisation is:
\begin{enumerate}
\item The dynamic variables $\E$, $m$, $h$, $u_a$, $w$, $o_a$,
$d$ are  `fast variables', \ie\ they change significantly during
the upstroke of a typical AP potential, unlike all other variables
which change only slightly during that period. 
The relative speed of the dynamical variables is estimated by
comparing the magnitude of their corresponding 'time scale
functions' as shown in \fig{f:0000}(a). For a system of
differential equations $\de\vec{y}/\de t  = \vec{F}(\vec{y})$ the
time scale functions are defined as $\tau_i(\vec{y}) \equiv \left| (
\de F_i /\de y_i )^{-1}\right|$, $i= 1\ldots N$ and coincide with the
functions $\tau$ already present in \eqs{e:CRN}.

\item A specific feature of $\E$ is that it is fast only because of one of
the terms in the right-hand side, the large current $\INa$, whereas
other currents are not that large and so do not have the large
coefficient $\epsilon^{-1}$  in front of them.
\item The fast sodium current $\INa$ is only large during the upstroke
of the AP, and not that large otherwise  as illustrated in
\fig{f:0000}(d). This is due to the fact that either
gate $m$ or gate $h$ or both are almost closed outside the upstroke 
since their quasistationary values $\mbar(\E)$ and $\hbar(\E)$ are small
there as seen in \fig{f:0000}(b). Thus
in the limit $\epsilon\to0$, functions $\mbar(\E)$ and $\hbar(\E)$ have to be
considered zero in  certain overlapping intervals $\E\in(-\infty,\Em]$
and $\E\in[\Eh,+\infty)$, and $\Eh\le\Em$, hence the representations
$\emb{\mbar}(\E;0)=M(\E)\,\Heav(\E-\Em)$ and
$\emb{\hbar}(\E;0)=H(\E)\,\Heav(\Eh-\E)$. 
\item 
The term $\Curv \D_{X}\E$ in the first equation represents the effect of the 
front curvature for waves propagating in two or three spatial dimensions.
Derivation of this term using asymptotic arguments can be found e.g. in
\cite{Tyson-Keener-1988}.
A simple rule-of-thumb way to understand it is this. Imagine
a circular wave in two spatial dimensions. The diffusion term in the
equation for $\E$ then has the form 
$D \left(\D_{X}^2 + \D_{Y}^2 \right){\E}
=D \left(\D_{R}^2 + \frac{1}{R} \D_{R} \right){\E}$
where $R$ is the polar radius. If $R$ at the front is large,
its instant curvature $\Curv=1/R$ changes slowly as the front propagates,
and can be replaced with a constant for long time intervals. 
Considering $R$ as a new $X$ coordinate, we then  get \eqs{e:CRN}.
\end{enumerate}
These aspects, as applied to the fast sodium current, have been shown
to be crucial for the correct description of the propagation
block \cite{Biktashev-2002}.
  In particular, it is important that the $h$-gate is included among the fast variables.
  The particular importance of $h$ dynamics at the fringe of excitability
  has been noted before, e.g. for the modified
  Beeler-Reuter model~\cite{Vinet-Roberge-1994}.
A more detailed discussion of the parameterisation
given by \eqs{e:CRN} can be found in \cite{Biktasheva-etal-2005}.

A change of variables\footnote{
  A change of the value of $D$ is equivalent to rescaling of the
  spatial coordinate, and is not critical to any of the questions
  considered here. In order to operate with dimensional velocity, we
  assume the value of the diffusion coefficient 
  $D=0.03125\,\mm^2/\ms$, 
  as in our earlier publications
  \cite{Biktasheva-etal-2003,Biktasheva-etal-2005}.
    Increase of the diffusion coefficient
    to, say, $D=0.1\,\mm^2/\ms$
    raises the propagation velocity from $0.28\,\mm/\ms$ in 
    Table~\ref{t:01} to $0.50\,\mm/\ms$, in full agreement
    e.g. with results of \citet{Xie-etal-2002} for the same model.
}
${t}=\epsilon^{-1}T$, ${x}=(\epsilon D)^{-1/2}X$,
${\curv}=(\epsilon D)^{1/2}\Curv$ and subsequently the limit
$\epsilon\to0$ transforms \eqs{e:CRN} into
\begin{align}
\hspace*{-6mm}
&\D_{{t}}{\E} = \left(\D_{{x}}^2 +\curv\D_{{x}}\right){\E} - \CM^{-1} \INa(\E,m,h,j),  \nonumber\\
&\D_{{t}}{m} =
{\big(M(\E)\,\Heav(\E-\Em)-m\big)}/{\tau_{m}(\E)},        \nonumber\\
&\D_{{t}}{h} =   {\big(H(\E)\,\Heav(\Eh-\E)-h\big)}/{\tau_{h}(\E)},        \nonumber\\
&\D_{{t}}{\ua} = {\big(\uabar(\E)-\ua\big)}/{\tau_{\ua}(\E)},    \nonumber\\
&\D_{{t}}{w} =   {\big(\wbar(\E)-w\big)}/{\tau_{w}(\E)},         \nonumber\\
&\D_{{t}}{\oa} = {\big(\oabar(\E)-\oa\big)}/{\tau_{\oa}(\E)},    \nonumber\\
&\D_{{t}}{d} =   {\big(\dbar(\E)-d\big)}/{\tau_{d}(\E)},        \nonumber\\
&\D_{{t}}{\U} =  0 .                                     \label{e:CRN0}
\end{align}
In other words, we consider the fast time scale on which the upstroke
of the AP happens, neglect the variations of slow variables
during this period as well as all transmembrane currents except $\INa$,
as they do not make significant contribution during this period and
replace $\mbar$ and $\hbar$ with zero when they are small.

In the resulting  \eqs{e:CRN0} the first three equations for
$\E$, $m$ and $h$ form a closed subsystem, the following four equations for
$\ua$, $w$, $\oa$ and $d$, can be solved if $\E(x,t)$ is known but do
not affect its dynamics, and the rest of the equations state that all
other variables remain unchanged.
Hence we concentrate on the first three equations  as the system
describing propagation of an AP front or its failure. The above
derivation procedure 
does not give a precise definition of the functions $H(\E)$ and $M(\E)$, it only 
requires that these are reasonably close to $\hbar(\E)$ and $\mbar(\E)$ for those
values of $\E$ where these functions are not small. Here `reasonably close'
means that replacement of 
$\hbar(\E)$ with $H(\E)\,\Heav(\Eh-\E)$ and
$\mbar(\E)$ with $M(\E)\,\Heav(\E-\Em)$ does not change significantly
the solutions of interest, \ie\ the propagating fronts. We have found
that the simplest approximation in the form $M(\E)=1$, $H(\E)=1$ works
well enough. This is demonstrated in table \ref{t:01} where various
choices of $M(\E)$ and $H(\E)$ are tested. So, ultimately, we consider 
the following system
\begin{subequations}
\label{e:0020}
\begin{align}
\label{e:0021}
& \dd_t \E = \left(\dd_x^2 +\curv \dd_x\right)\E + \INabar(\E)\,j\,h\,m^3, \\
\label{e:0022}
& \dd_t h = \big(\Heav(\Eh-\E) - h\big)/\tau_h(\E), \\
\label{e:0023}
& \dd_t m = \big(\Heav(\E-\Em) - m\big)/\tau_m(\E),
\end{align}
\end{subequations}
where 
\begin{subequations}
\label{e:0030a}
\begin{align}
& \label{e:0030.1}
\INabar(\E) = g_{Na}(\ENa - \E), \\
&  \label{e:0030.2}
\tau_k(\E) = \big(\alpha_k(\E)+\beta_k(\E)\big)^{-1}, \qquad k=h, m,  \\
& \alpha_h(\E) = 0.135\,e^{-(\E+80)/6.8}\,\Heav(-\E-40), \nonumber \\
& \beta_h(\E) = \left(3.56\,e^{0.079\E}+3.1\times 10^5\,e^{0.35\E}\right)\,
\Heav(-\E-40) \nonumber \\
&\hspace{2cm} + \Heav(\E+40)\,\big(0.13 (1+e^{-(\E+10.66)/11.1})\big)^{-1}, \nonumber \\
& \alpha_m(\E) = \f{0.32 (\E+47.13)}{1-e^{-0.1(\E+47.13)}}, \nonumber \\
& \beta_m(\E) = 0.08 e^{-\E/11}, \nonumber \\
& g_{Na}=7.8, \quad \ENa=67.53,  \quad \Eh=-66.66, \quad \Em=-32.7. \nonumber
\end{align}
\end{subequations}
All parameters and functions here are defined as in \cite{CRN98} except the
new `gate threshold' parameters $\Eh$ and $\Em$ which are
chosen from the conditions $\hbar(\Eh)=1/2$ and $\mbar^3(\Em)=1/2$.  
As follows from the derivation, variable $j$, the slow inactivation gate
of the fast sodium current, acts as a parameter of the
model. It is the only one of all slow variables included in the vector $\U$ 
that affects our
fast subsystem. We say that it describes the `excitability' of the
tissue. Notice that it is a multiplier of $\gNa$, so a reduced
availability of the fast sodium channels, \eg\ as under 
tetrodotoxin~\cite{Matthews-2003} or arguably in Brugada
syndrome~\cite{Antzelevitch-etal-2005} can be \reva{formally}
described by a reduced value of the parameter $j$.

Before proceeding to the analysis of the simplified three-variable
model defined by \eqs{e:0020} we wish to demonstrate that it is a good
approximation of the full model of \cite{CRN98} both  on a qualitative
and a  quantitative level.  
On the qualitative level, we show that a temporary obstacle
leads to a dissipation of the front. This is illustrated
in \fig{f:0010} which  
shows propagation of the AP
into a region in time and space where the excitability of
the tissue is artificially suppressed. The sharp wave fronts of
the model of \citet{CRN98} as well as of \eqs{e:0020}
stop propagating and start to spread diffusively once they reach the
blocked zone. The 
propagation does not resume after the block is removed. This behaviour
is completely different from that of the FitzHugh-Nagumo system of
\eqs{FHN} in which even though the propagation is blocked for
nearly the whole duration of the AP, the wave
resumes once the block is removed. 
Table~\ref{t:01} illustrates, on the quantitative level,
the accuracy of \eqs{e:0020} as an approximation of the full model of
\cite{CRN98}.  

It is a popular concept going back to classical works
\cite[\eg{}][]{Krinsky-Kokoz-1973-3} that the fast activation gate $m$
is considered a `fast variable' and is `adiabatically eliminated' since most
of the time, except possibly during a very short transient, it is close to
its quasistationary value $m\approx \mbar(\E)$. Hence the model can
be simplified by replacing $m$ with $\mbar(\E)$ and eliminating the
equation for $m$, 
\begin{align}
\label{e:0020a}
& \dd_t \E = \dd_x^2 \E + \INabar\,\Heav(\E-\Em)\,j\,h, \nonumber \\
& \dd_t h = \big(\Heav(\Eh-\E) - h\big)/\tau_h.                  
\end{align}
We have explored this possibility for the model of \citet{CRN98} in  \cite{Biktasheva-etal-2005}.
\eqs{e:0020a} are qualitatively correct, \ie{} they still show front
dissipation on collision with  a temporary obstacle, but make a large
error in the  front propagation speed, as demonstrated in table~\ref{t:01}.

\subsection{Travelling waves and reduction to ODE of the  three-variable model}
\label{ssec:2.3}

To find out when propagation of excitation is possible in our simplified model
and when it will be blocked, we study solutions in the form of
propagating fronts as well as the conditions of existence of such solutions.

We look for solutions in the form of a front propagating
with a constant speed and shape. So we use
the ansatz $F(z) = F(x+ct)$ for $F=\E,h, m$ 
where $z=x+ct$ is a `travelling wave coordinate' and $c$ is the 
dimensionless wave
speed of the front, 
related to the dimensional speed $C$ by 
$c=(\epsilon/D)^{1/2}C$.
Then \eqs{e:0020} reduce to a system of
autonomous ordinary differential equations,  
\begin{subequations}
\label{e:0030}
\begin{align}
\label{e:0031}
& \E'' = (c - \curv)\, \E' - \INabar(\E)\,j\,h\,m^3, \\
\label{e:0032}
& h' =  \big(c\,\tau_h(\E) \big)^{-1} \big(\Heav(\Eh-\E) - h \big), \\
\label{e:0033}
& m' =  \big(c\,\tau_m(\E) \big)^{-1}  \big(\Heav(\E-\Em) - m \big),
\end{align}
\end{subequations}
where the boundary conditions are given by 
\begin{subequations}
\label{e:0040}
\begin{align}
& \E(-\infty) = \Ealpha, & & \E(+\infty) = \Eomega, \quad  \Ealpha < \Eh < \Em < \Eomega, \\
& h(-\infty)=1, & & h(+\infty)=0,  \\
& m(-\infty)=0, & & m(+\infty)=1. 
\end{align}
\end{subequations}
Here $\Ealpha$ and $\Eomega$ are the pre- and post-front voltages.  

\eqs{e:0030} represent a system of fourth order so its general solution 
depends on four arbitrary constants. Together with constants
$\Ealpha$, $\Eomega$ and $c$ this makes seven constants to be determined
from the six boundary conditions in \eqs{e:0040}. Thus, we should have
a one-parameter family of solutions, \ie{} one of the parameters
$(\Ealpha, \Eomega, c)$ can be chosen arbitrary from a certain range.
A natural choice is $\Ealpha$ because the pre-front voltage acts as an
initial condition for a propagating front in the tissue, and because in our
study of the conditions for propagation it is most conveniently treated
as a parameter rather than as an unknown.

\section{Analytical study of the reduced model}
\label{sec:3}

\subsection{An exactly solvable caricature model}

The parameter-counting arguments given in the previous section make it 
plausible that the problem defined by \eqs{e:0030} with boundary
conditions of
\eqs{e:0040} has a one-parameter family of travelling wave-front
solutions. However, the problem is posed in a highly unusual way since
the asymptotic pre-front and post-front states are not stable
isolated equilibria but belong to continua of equilibria and thus are
only neutrally stable. We are not aware of any 
general theorems that would guarantee
existence of solutions of a nonlinear boundary value-eigenvalue problem
of this kind. For the two-component model of \eqs{e:0020a} considered in 
\cite{Biktasheva-etal-2005} this worry has been alleviated by the fact
that there is a `caricature' model which has the same structure
as \eqs{e:0020a} including the structure and stability of the
equilibrium set and which admits an exact and exhaustive analytical
study~\cite{Biktashev-2002}. 
Fortunately, a similar `caricature' exists for our present
three-variable problem as well.
We replace functions ${\INabar(\E)}$, ${\tau_h(\E)}$ and
${\tau_m(\E)}$ defined in \eqs{e:0030a} with constants.  The choice of
the constants is somewhat arbitrary. We assume that the events in the
beginning of the interval $z\in[\xi,+\infty)$, where $\E$ is just
above $\Em$, are most important for the front propagation. So for
numerical illustrations we choose the values of constants ${\INabar}$,
${\tau_h}$ and ${\tau_m}$ as the values of the corresponding functions
in \eqs{e:0030a} at some fixed value of the voltage $\E$.
We set the $z$ axis so that $\E(0)=\Eh$, 
and then $\E(\xi)=\Em$ for some $\xi>0$ still to be determined.
We demand that the solutions for
the unknowns $\E$, $h$ and $m$ are continuous and that $\E$
is smooth at the internal boundary points.

In this formulation, \eqs{e:0032} and \ref{e:0033} decouple
from \eq{e:0030}a and from each other and solved
separately. The solutions of these first-order linear ODE with
constant coefficients are given by \eqs{e:0050}b and \ref{e:0050}c,
respectively. It follows that in the interval $\E \leq \Em$,
\eq{e:0031} is a linear homogeneous ODE with constant coefficients and
its solution given at the first row of \eq{e:0050}a satisfies the
boundary conditions $\E(-\infty)=\Ealpha$, $\E(0)=\Eh$ and
$\E(\xi)=\Em$ provided that the internal boundary point $\xi$ is given
by \eq{e:0040:i0030}. To solve the linear inhomogeneous \eq{e:0050} in the
interval $\E \geq \Em$ we note that its inhomogeneous term $f\equiv
\INabar(\E)\,j\,h\,m^3$ is a sum of exponentials
\begin{align}
\label{e.0045}
& \hspace*{-0.3cm}
f = \INabar(\E)\,j\ \sum_{n=0}^3 (-1)^n \dbinom{3}{n}
e^{n\xi/(c\tau_m)} e^{-B_n z/c}, \\
& \hspace*{-0.3cm}
B_n \equiv \f{1}{\tau_h} + \f{n}{\tau_m} = \f{\tau_m+ n\,\tau_h}{\tau_h\, \tau_m}, \nonumber
\end{align}
and terms proportional to $n \tau_h$ will appear in the solution
due to the expression for $B_n$. Imposing the boundary conditions at
the internal point  
$\E(\xi)=\Em$ and at infinity $\E(\infty)=\Eomega$, we obtain the
solution in this interval given at the second row of \eq{e:0050}a.
Finally, the wave speed $c$ is fixed by \eq{e.0062} from the
requirement that the solution for $\E(z)$ is smooth at the internal
boundary point $\xi$.  To summarise, the solution of \eqs{e:0030} and
\ref{e:0040} is 
\begin{subequations}
\label{e:0050}
\begin{align}
&\E(z) = \hspace{.3mm}\left\{\begin{array}{l}
    (\Eh-\Ealpha)\,  e^{(c-\curv)z} + \Ealpha,  \\[1ex]
    \displaystyle \Eomega-\INabar\, j\, c^2 \tau_h^2 \tau_m^2 \sum_{n=0}^3 A_n(c,z),      
  \end{array}\right. 
  &
  \begin{array}{l}
    \vphantom{(\Eh-\Ealpha)\,  e^{(c-\curv)z} + \Ealpha,}
     z \leq \xi,\\ [1ex]
    \vphantom{\displaystyle \Eomega-\INabar\, j\, c^2 \tau_h^2 \tau_m^2 \sum_{n=0}^3 A_n(c,z),}
     z \geq \xi,
  \end{array}
\\[1ex]
&h(z) = \hspace{1.3mm} \left\{\begin{array}{l}
    1, \\[1ex]
    e^{-z/(c\,\tau_h)},
  \end{array}\right. 
  &
  \begin{array}{l}
    z \leq 0, \\[1ex]
    z \geq 0,
  \end{array}
\\[1ex]
&m(z) = \left\{\begin{array}{lr}
    0, \\[1ex]
    1-e^{(\xi-z)/(c\,\tau_m)},
  \end{array}\right. 
  &
  \begin{array}{l}
    z \leq \xi, \\ [1ex]
    z \geq \xi,
  \end{array}
\end{align}
\end{subequations}
where the pre-front voltage $\Ealpha$,
the post-front voltage $\Eomega$ and the  wave speed $c$ are
related by
\begin{subequations}
\label{e.0060}
\begin{align}
\label{e.0061}
& \hspace*{-0.3cm}
\Eomega = \Em+ \INabar\,j\, (c\, \tau_h\, \tau_m)^2 e^{-\xi/(c\,\tau_h)} \sum_{n=0}^3 \f{a_n(c)}{\tau_m+n\,\tau_h}, \\
\label{e.0062}
& \hspace*{-0.3cm}
 0 = (c-\curv)(\Em - \Ealpha) - \INabar\,j\,c\,\tau_h \tau_m\, e^{-\xi/(c\tau_h)}\sum_{n=0}^3 a_n(c), 
\end{align}
\end{subequations}
the distance between points $\E=\Eh$ and $\E=\Em$ is 
\begin{equation}
\label{e:0040:i0030}
\xi = \f{1}{(c-\curv)}\ln\!\left(\f{\Em-\Ealpha}{\Eh-\Ealpha}\right),
\end{equation}
and $A_n(c,z)$ and $a_n(c)$ are abbreviations for
\begin{subequations} \label{An.an}
\begin{align}
& A_n(c,z) \oth{\equiv} \f{a_n(c)}{\tau_m+n\tau_h} \, 
        \exp\left( \f{n\xi\tau_h - (\tau_m+n \tau_h) z}{c\,\tau_h\,\tau_m} \right), \label{An} \\
& a_n(c) \oth{\equiv} \dbinom{3}{n}\f{(-1)^n}{c(c-\curv)\, \tau_h\, \tau_m+\tau_m+n\, \tau_h}. \label{an}
\end{align}
\end{subequations}
 In the limit $\tau_m\to0$ this solution  
tends to the solution of the two-component model of \cite{Biktashev-2003}, as expected.

The accurate expression in \eq{e:0030.1} for the
sodium current $\INabar(\E)$ vanishes for $\E=\ENa$ 
which, in particular, means that the transmembrane voltage never
exceeds $\ENa$. So, replacing this function with a constant changes the
properties of the system qualitatively. Even bigger 
discrepancies are expected to occur from replacing the ${\tau_h(\E)}$ and
${\tau_m(\E)}$ by constants because these functions vary by an order of
magnitude in the range between the pre- and the post-front voltage.
It is surprising, however, that even this rough approximation produces
results which, with exception of the post-front voltage, are within
several percent from the solution of the detailed ionic model
\cite{CRN98} and certainly capture its qualitative features as
can be seen in \fig{f:0020} where the constants are chosen at
  $\E=\Em$, \ie{} $\INabar(\Em)$, ${\tau_h(\Em)}$ and ${\tau_m(\Em)}$.
This relatively good agreement is not due to this special choice of
parameter values. Indeed, the caricature model 
and its solution \eqs{e:0050} involve the parameters  $\INabar$,
$\tau_h$, $\tau_m$, $\curv$, $\Ealpha$ and $j$. The dependence on the
curvature $\curv$ is negligible in comparison to the deviation of the
solution \eqs{e:0050} of the caricature model from the numerical
solution of the three-variable model \eqs{e:0030}. The dependence on the
pre-front voltage $\Ealpha$ and the excitability parameter $j$ is
discussed in section \ref{ssec:3.2} and represented in Figs. \ref{f:0030} and
\ref{f:0040}. The 
parameters $\INabar$, $\tau_h$, $\tau_m$, on the other hand, are
somewhat arbitrary but in order to achieve a good agreement with 
the original system given by \eqs{e:0030} we choose these
values as the values of the corresponding functions in \eqs{e:0030a}
at various values of $\E$. In \fig{f:0030} the relationship between the wave
speed $c$ and the excitation parameter $j$ for several such choices of
$\E$ is presented. It can be seen that such a variation of the values
of  $\INabar$, $\tau_h$, $\tau_m$ does not lead to significant
qualitative changes in the solution   \eqs{e:0050} of the caricature model.
Figs.~\ref{f:0020} and \ref{f:0030} also show, for comparison, 
the numerical solutions of the detailed ionic model of \cite{CRN98}
and of the full three-variable model of \eqs{e:0030}, which will be
described in detail in the next section.

\subsection{The condition for propagation}
\label{ssec:3.2}

\Eq{e.0062} defines $c$ as a smooth function of the parameters
within a certain domain.
The boundary of this domain is associated with the propagation failure.
Not all parameters, $\INabar$, $\tau_h$, $\tau_m$, $\curv$, $\Ealpha$ and $j$,
entering  \eq{e.0062} are of equal importance. 
We consider here
$\curv = 0$ and postpone the
investigation of the effects of curvature to the next section.
Parameters $\INabar$, $\tau_h$ and $\tau_m$
represent
well-defined 
properties of the tissue, albeit changeable depending on physiological conditions.
On the other hand,
parameters $j$ and $\Ealpha$ are not model constants, but `slowly varying'
dynamic quantities: $j$ 
remains approximately constant throughout the front,
and $\Ealpha$ represents the transmembrane voltage ahead of the front,
but both can vary widely on large scales between different fronts. 
Hence we need to determine the singular points of the
dispersion relation in \eq{e.0062} with respect to $j$ and $\Ealpha$.

Similarly to the two-component caricature \cite{Biktashev-2002},
\eq{e.0062} is a transcendental equation for $c$,
but it is easily
solvable for the excitation parameter $j$:
\begin{equation}
\label{e:0070}
j= \f{(\Em-\Ealpha)}{6\, \INabar\, \tau_h^{4}\, \tau_m}\,
e^{\f{\xi}{c\, \tau_h}} \prod_{n=0}^3 
\big(c^2\, \tau_h\, \tau_m+\tau_m+n\, \tau_h\big).
\end{equation}
The resulting relationship of $j$ and $c$ 
for a selected value of $\Ealpha$
is
shown in \fig{f:0030}. This figure 
reveals a bifurcation.
For
values of $j$  lower than some $\jmin$ no travelling
wave solutions exist. After a bifurcation at $j > \jmin$ two
solutions with different speeds are possible. Our direct numerical
simulations of \eqs{e:0020} as well as studies of the two-component 
caricature
model by \citet{Hinch-2004} suggest that the solutions of the lower
branch are unstable. 
The bifurcation point $\jmin$ can be determined
from the 
condition
that $j$ has a minimum with respect to $c$
at this point and therefore satisfies
\begin{equation}
\label{e:0080}
\left(\f{\dd j}{\dd c}\right)_{\Ealpha=\mbox{const}} = 0. 
\end{equation}
This produces, with $j(c)$ defined by \eq{e:0070},
a
quintic polynomial equation for $c^2$. 

Activation of the sodium current is possible because $\tau_m \ll
\tau_h$, permitting transient channel opening and
current flow through the cell membrane. The ratio $\tau_h/\tau_m$
is a function of $\E$ in the full model, and is a constant in
\eqs{e:0030}. The minimal value of this ratio, necessary
for propagation, is shown on \fig{f:0031} as a function of
various choices of $\INabar$, $\tau_m$ and $j$; it is 
obtained by numerical solution of the algebraic equation \eq{e.0062}.
The smallness of $\tau_m/\tau_h$ allows approximate solution of
the above mentioned quintic equation for $c^2$. We set
\begin{equation}
\label{e:0090}
c^2 = \sum_{n=0}^\infty S_n \tau_m^{n} .
\end{equation}
Substituting this expansion in \eq{e:0080} and discarding the small
terms of order $O(\tau_m)$ gives the zeroth-order approximation to the
solution as a function of the pre-front voltage $\Ealpha$
\begin{align}
\label{e:0100}
&\jmin^{(0)} = 
  \frac{(\Em-\Ealpha)}{2 \INabar \tau_h} 
  e^{ \frac{
    2\Theta
  }{
    \Theta+\sqrt{\Theta^2+4\Theta}
  } } \,  
  \left(\Theta+2+\sqrt{\Theta^2+4\Theta}\right), 
  \\
&
\Theta = \ln\!\big( (\Em-\Ealpha)/(\Eh-\Ealpha) \big).  \nonumber
\end{align}
This limit corresponds to the two-variable caricature
\cite{Biktashev-2002}.  
\textit{For any given value of the pre-front voltage the value 
of $j$ must be larger than $\jmin$ in order for wave fronts to
propagate.} Although lacking sufficient
accuracy, the zeroth-order approximation given by
\eq{e:0100} reproduces qualitatively well the
conditions for propagation and dissipation of excitation fronts in
the model of \citet{CRN98}.
Analogously, discarding the small terms of order $O(\tau_m^2)$
gives the first-order approximation, 
\begin{align}
\label{e:0101}
&\jmin^{(1)} = \f{(\E_m-\Ealpha)}{6\,\Delta^4\,\tau_m}\,
  e^{-\frac{\Delta\Theta}{(A\tau_h-\Delta\Theta)}}
\prod_{n=0}^{3}(A\,\tau_m-n\,\Delta), \\ 
& \Delta = 12\, \tau_h^2\sqrt{\Theta\,(\Theta+4)}, \nonumber \\
&  A= \left(
    \Theta^2 (\Theta + 4) + 
    \Theta^{3/2} (\Theta + 2) \sqrt{\Theta+4} 
  \right) \left(11\,\tau_m-6\,\f{\tau_h}{\Theta} \right) . \nonumber
\end{align}
This approximation is already very good and changes insignificantly as
more terms are considered in  \eq{e:0090},  see \fig{f:0040}.


\section{Numerical results}
\label{sec:4}

\subsection{Propagating front solutions}

We solved \eqs{e:0030}--\ref{e:0040} numerically,
using the method described in Appendix \ref{appx:1}.
The results are shown in figures
\ref{f:0020}, \ref{f:0030}  and \ref{f:0050}. \Fig{f:0020} 
offers a comparison of the shapes of the solution of \eqs{e:0030} with
a snapshot of a travelling wave solution of the full model of
\citet{CRN98}. 
The values of the wave speed and the post-front voltage are
presented in table \ref{t:01}  and also show an excellent agreement.  
This confirms our assumptions that the fronts of travelling waves
in the full model have constant speed and shape and thus 
satisfy an ODE system, and that $j$ remains approximately constant
during the front. 
\Fig{f:0050} shows the wave speed $c$ as a function of two of the parameters
of the problem, the pre-front voltage $\Ealpha$ and the excitability
parameter $j$. 
For every value of $j$ and $\Ealpha$  from a certain domain,
two values of the wave speed $c$ are possible, 
which is similar to the solutions of the caricature model.
The smaller values of $c$ are not observed in the PDE simulation of the
full model. This is a strong indication that they are unstable.

\subsection{The condition for propagation}

In this subsection, we report numerical values for the threshold of
excitability $\jmin$ below which wave fronts are not sustainable and
have to dissipate,  as predicted by the reduced three-variable model
of \eqs{e:0030}--\ref{e:0040}.  \Fig{f:0060} presents $\jmin$ as  a
function of the pre-front voltage $\Ealpha$.  
  The curve $\jmin(\Ealpha)$ represents a boundary in the space
  of the slow variables $(\E,j)$ which separates the region of relative
  refractoriness where   excitation fronts are possible, even though
  possibly slowed down,   from the region of absolute refractoriness
  where excitation   fronts cannot propagate at all. 
  In practice, however, we can reduce the condition of the absolute
  refractoriness even further. This is possible because typical APs
  have their tails very closely following one path on the $(\E,j)$  plane.
  This property is known for cardiac models; e.g. \cite{Vinet-Roberge-1994}
  presents an evidence for the Modified Beeler-Reuter model
  that the dynamics of recovery from an AP do not depend on details
  of how that AP has been initiated.
  Therefore of the whole curve $(\E,\jmin(\E))$ only one point is important,
  its intersection with the curve $(\E(t),j(t))$ representing
  a typical AP tail. 
  For the model \citet{CRN98} considered here, we simply state
  the existence of this   universal $(\E(t),j(t))$ curve   as an
  "experimental fact".  This is illustrated in \fig{f:0060} where we
  plot   the curve $(\E,\jmin(\E))$ together with   projections of a
  selected set of AP   trajectories.  
The AP solutions were obtained for a space-clamped
version of \cite{CRN98} with initial conditions for $j$
and $\E$ as shown in the figure and all other variables in their
resting states. These trajectories allow us to follow the correlation
between the transient of $j$ and the AP $\E$. Indeed, in the  tail of an AP
solution, the curve $j$ vs $\E$ is almost independent of the way
the AP is initiated. As a result,
the projections of the trajectories $(\E(t),j(t))$
intersect the critical curve $(\Ealpha,\jmin(\Ealpha))$
in a small vicinity of one point,
$(\jcr,\Ecr)=(0.2975\pm0.0015,-72.5\pm0.5)$.
This result suggests the following interpretation. 
As a wave front propagating into the tail of a
preceding wave reaches a point in the state corresponding to  this
"absolute refractoriness" point $(\jcr,\Ecr)$, it will stop because of
insufficient excitability of the medium, and dissipate.

  In a broader context, in the front propagation speed $c$ is a
  function of $j$ and $\E$ in the relative refractoriness region of
  the $(\E,j)$ plane, so the highly correlated dependencies of $\E(t)$
  and $j(t)$ in the wake of an AP mean that $c$ at a particular point
  becomes a fixed function of time. This makes it possible to describe
  $c$ in terms of the diastolic interval (DI), i.e the time passed
  after the end of the preceding AP. This dependence, known as
  dispersion curve or velocity restitution curve, is an important tool
  in simplified analysis of complex regimes of excitation
  propagation~\cite{%
    Nolasco-Dahlen-1968,%
    Courtemanche-etal-1993,%
    Vinet-Roberge-1994a,%
    Wellner-Pertsov-1997,%
    Weiss-etal-1999%
  }.

\subsection{Propagation block in two dimensions}

In two spatial dimensions, the condition of dissipation $j<\jcr$
may happen to a piece of a 
wave front rather than the whole of it. In that case 
we observe a local block and a break-up of the excitation wave. 
\Fig{f:0070} shows how it happens in a 
two-dimensional simulation of the detailed model of \citet{CRN98}. A
spiral wave was initiated by 
a cross-field protocol.  This spiral wave develops instability, breaks
up from time to time, and eventually self-terminates. This is one of
the simulations discussed in detail in \cite{Biktasheva-etal-2005}.  Here
we use it to test our newly obtained criterion of propagation
block. The red colour component represents the $\E$ field, white for
the resting state and maximum for the AP peak. This is superimposed
onto an all-or-none representation of the $j$ field, with black for
$j>\jcr$ and blue for $j\le\jcr$. Thus the red rim represents the
`active front' zone where excitation has already happened but $j$
gates are not de-activated yet; most of the excited region is in shades of
purple representing the gradual decay of the AP with $j$ deactivated. The
wave ends up with a blue tail, which corresponds to $\E$ already
close to the resting potential but $j$ not yet recovered and still
below $\jcr$. So the blue zone is where there is no excitation, but
propagation of excitation wave is impossible, \ie\ absolutely
refractory zone. The black zone after the tail and before the new
front is therefore relative refractory zone, where front propagation
is possible. Thus, in terms of the colour coding of
figure~\ref{f:0070}, the prediction of the theory is: the wavefront
will be blocked and dissipate where and when it reaches the blue zone,
and only there and then.
This is exactly what happens in the shown panels: the red front touches 
the blue tail, first at the third panel, at the point indicated by the
white arrow, and subsequently in its vicinity. The excitation front stops
in that vicinity and dissipates.
So we have a break-up of the front.

The analysis of the numerics, which ran for the total of 7400\,ms 
until self-termination of the spiral and
showed 4 episodes of front break-up, has confirmed that in all cases
the break-up happened if and only if the front reached the blue
region $j\le\jcr$.

\subsection{Curvature effects}
\label{ss:4.3}

Since we attempt to compare the results of our one-dimensional model to
simulations of spiral waves in two-dimensions, it is important to
explore the dependence of the solution on the curvature of the front.
The standard theory says
that in two dimensions the normal velocity of the wave
front need to be corrected by the term $\lambda \Curv$ where $\lambda$ is the typical
width of the wave front \cite{Tyson-Keener-1988}. 
The speed-curvature
diagram presented in \fig{f:0080}(a) shows that 
in our simplified model
this relationship is
satisfied to rather large values of $|\Curv|$. 
Our choice of boundary conditions in \eqs{e:0040} 
assumes that
the excitation fronts propagate
from right to left,
so 
positive values of the curvature
correspond to concave 
fronts.
Only
at very small values of the radius of curvature of the order of 0.3 mm
for $j=1$ the wave speed shows a non-linear dependence on curvature as
seen in the insert \fig{f:0080}(b). This part of the figure also
demonstrates that there is a critical value of the curvature for which
the excitation wave stops to propagate as well as an unstable branch
of the solution. However, these phenomena occur at very large
curvatures which are far outside of the range of values 
of $|\Curv|<0.1\mm^{-1}$ observed in the two-dimensional simulations
of \fig{f:0070}.

The most important question with respect to our study is whether the
curvature  changes significantly the critical value of the excitation
parameter $\jcr$  below which the wave fronts fail to propagate.  To
answer this question we present \fig{f:0080}(c) in which  the wave
speed $c$ is shown as a function of $j$ for three values of the
curvature corresponding to a non-curved front and to convex and
concave fronts with radius of curvature equal to 10 mm. The values of
$\jmin$ for these three cases differ only slightly.
So, the propagation blocks in our simulations do not
depend significantly on the curvature of the front.

  This conclusion is valid for the particular
  cardiac model~\cite{CRN98} and for the particular
  context. In \cite{Comtois-Vinet-1999}, the minimal diastolic interval,
  defined as time from the moment $\E=-50$\,mV to the moment propagation
  becomes possible again, depended only slightly on curvature
  for the Modified Beeler-Reuter model at standard parameters, 
  but was much more pronounced when $\tau_j$ was artificially
  increased 6-fold. The simplest explanation of this difference is
  that the small variation of $\jmin$ due to the curvature takes
  much longer for $j(t)$ to make if $\D{j}/\D{t}$ is very small,
  so even that small variation $\jmin$ becomes significant.

\section{Conclusions}
\label{sec:5}

In this paper, we have shown that propagation of excitation and its block 
in \citet{CRN98} model of
human atrial tissue can be successfully predicted by a simplified
model of the excitation front, obtained by 
an asymptotic
description focussed on the fast sodium current, $\INabar$.
Whereas it was known that main qualitative features of the
$\INa$-driven fronts can be described by a two-component model for
$\E$ and $h$, we have now found that for good quantitative predictions,
one must also take into account the dynamics of $m$ gates. 
Thus, we have proposed a three-component description of the
propagating excitation fronts given by \eqs{e:0020}.  
We have obtained an exact analytical solution for piecewise-linear
`caricature' three-component model of \eqs{e:0020}. For an appropriate 
choice of parameters, it reproduces the key qualitative features of
the accurate three-component model of \eqs{e:0020} and gives a correct
order of magnitude quantitatively.
Numerical solution of the automodel equation of the proposed
three-component model of \eqs{e:0020} gives a very accurate 
prediction of propagation block in two-dimensional re-entrant
waves.  For the given model, this reduces to a condition involving the
pre-front values of $\E$ and $j$, or even in terms of $j$ alone. This
provides the sought-for operational definition of absolute
refractoriness in terms of $j$, simple and efficient.

The success of the propagation block prediction justifies the
assumptions made on the asymptotic structure, \ie{} appearance of the
small parameter $\epsilon$, of \eqs{e:CRN}, and also confirms that
two-dimensional effects, \eg{} front curvature, do not significantly
affect the propagation block conditions, at least in the particular
simulation. 

As the description and role of $\INa$ are fairly universal in
cardiac models, most of the results should be applicable to other
models. However, some other cardiac models may require a more
complicated description. For instance, the contemporary `Markovian'
description  of $\INa$ \cite[][\eg\ ]{Clancy-Rudy-2001} is very
different from the classical $m^3jh$ scheme.  Also, propagation in
ventricular tissue in certain circumstances can be essentially
supported by  L-type calcium current rather than mostly $\INa$ alone
\cite{Shaw-Rudy-1997}.  

{\small

\appendix
\section{Numerical method}
\label{appx:1}

For a numerical solution the problem needs to be formulated on a
finite interval $z \in [z_{min}, z_{max}]$ rather than on the open
interval $z \in (-\infty, \infty)$. 
Furthermore, because of the piecewise definition of the problem this
interval must be separated in three parts   $[z_{min}, 0]$, $[0, \xi]$
and $[\xi, z_{max}]$ as discussed in the beginning of section
\ref{sec:3}. The standard numerical methods we use require that the
problem is posed on a single interval,  for instance $y \in [0, L]$.
So we use the mapping, 
\begin{align}
\label{e:0120}
&  [0,L] \ni y =  \left\{
\begin{array}{lll}
 -z, & z \in [z_{min},0], \\ 
 (\xi/L)\, z, &  z \in [0,\xi],  \\ 
 z - \xi, \qquad & z \in [\xi, z_{max}], 
\end{array}\right.
\end{align}
to transform \eqs{e:0030} as follows
\begin{align}
\label{e:0130}
& \E_1'' = -(c-\curv)\, \E_1' + \gNa (\ENa-\E_1)\,j\,h_1\,m_1^3, \nonumber \\
& h_1' =  -\big(c\,\tau_h(\E_1) \big)^{-1} \big(1 - h_1 \big), \nonumber \\
& m_1' =  \big(c\,\tau_m(\E_1) \big)^{-1}  m_1,  \nonumber \\
& \E_2'' = \big((c-\curv)\, \E_2' - \gNa (\ENa-\E_2)\,j\,h_2\,m_2^3\big)/p, \nonumber \\
& h_2' =  -\big(p\,c\,\tau_h(\E_2) \big)^{-1} h_2, \nonumber \\ & m_2' =  -\big(p\,c\,\tau_m(\E_2) \big)^{-1} m_2,   \\
& \E_3'' = (c-\curv)\, \E_3' - \gNa (\ENa-\E_3)\,j\,h_3\,m_3^3, \nonumber \\
& h_3' =  -\big(c\,\tau_h(\E_3) \big)^{-1} h_3, \nonumber \\
& m_3' =  \big(c\,\tau_m(\E_3) \big)^{-1} (1-m_3),  \nonumber \\
& c' = 0,\nonumber \\
& p'=0, \qquad \qquad \mbox{where $p\equiv \xi/L$}\nonumber \\
& \Eomega'=0, \nonumber
\end{align}
where the subscripts 1, 2 and 3 denote the variables
corresponding to the three subintervals.
Here, the end of the second subinterval $\xi$ is an unknown parameter
and together with the wave speed $c$ and the post-front voltage
$\Eomega$ must be determined as a part of the solution.  Because these
unknowns are constants, their derivatives must vanish which leads to
the introduction of the last three equations in \eqs{e:0130}.

The boundary conditions in \eqs{e:0040} 
at infinity 
are substituted by 
\begin{equation}
\label{e:0140}
\Big(\vec u \Big)_{z_{min}, z_{max}} = \Big(\vec u \Big)_{(\mp\infty)}
+ \vec v,
\end{equation}
where $\vec u$ is the vector of unknown variables and $\vec v$ is a
vector of small perturbations,
obtained as a solution of
\eqs{e:0030} linearised about \eqs{e:0040}. Together
with the implicit assumptions   $\E(0)=\Em$ and $\E(\xi)=\Eh$ which
break the translational invariance and
the additional  requirements that the solutions must be continuous
functions of $z$ and that $\E(z)$ must be smooth, the necessary 15
conditions are 
\begin{align}
\label{e:0150}
& \E_1(0)=\E_h, \quad \E_2(0) =\E_h, \quad \E_3(0) =\E_m, \nonumber \\
& \E_1'(0)=-p(0)\,\E_2'(0), \quad h_1(0)=h_2(0), \quad m_1(0)=m_2(0),\nonumber \\
& \E_3'(0)=p(L)\,\E_2'(L), \quad h_3(0)=h_2(L), \quad
m_3(0)=m_2(L), \nonumber\\
& \E_1'(L)=-(c(L)-\curv)\,\big(\E_1(L)+\Ealpha\big), \quad \E_2(L)=\E_m,  \\
& \E_3(L) =-\big(\E_3(L)-\Eomega(L)\big)/\left(c(L)\,\tau_h\big(\E_3(L)\big)\right),  \nonumber \\
& h_1(L)=1, \quad  m_1(L)=0, \nonumber\\
& h_3(L) = \f{\E_3'(L)}{\gNa\,j\,\big(\ENa-\E_3(L)\big)}\,\left(
\f{1}{c(L)\, \tau_h\big(\E_3(L)\big)} + (c(L)-\curv)\right).\nonumber
\end{align}
We use the boundary-value problem solver {\tt D02RAF} of the NAG
numerical library which employs a finite-difference discretization
coupled to a deferred correction
technique and Newton iteration \cite{nag_fl_21}. The analytical
solution given in \eqs{e:0050} is used as an initial approximation to start
the correction process. The method proves to be very robust over a
large range of parameters.  

The authors are grateful to I.V.~Biktasheva for sharing her experience of
simulation of model \cite{CRN98}, to H.~Zhang and P.~Hunter for inspiring
discussions related to this manuscript \oth{and to the anonimous
referees for constructive criticism and helpful suggestions}. This work is supported by
EPSRC grants 
GR/S43498/01 
and
GR/S75314/01. 
}


\clearpage
\newcommand{\mytable}[3]{
\begin{table}[p]
#1
\caption[]{#2}
\label{#3}
\end{table}
}                                                               \typeout{tabs started}
\mytable{
\centerline{\begin{tabular}%
{|l|c|c|c|c|}\hline
{\bf Model}  & {\bf Wave speed}   &  {\bf Rel. error}  & {\bf Post-front}   &
                        \reva{{\bf Maximum rate}} \\
             &          &      & {\bf voltage} & {\bf of AP rise} \\
             & $C$, [mm/ms]       &  in $C$            &  $\Eomega$, [mV] 
             &   $(\de V / \de t)_\mathrm{max}$ [V$/$s] \\[1mm]\hline
The full model of       & & & & \\[1mm]
\citet{CRN98}
                        & 0.2824      & --            &     3.60   & 173.83   \\[1mm]\hline
Model \cite{CRN98} with replacements & & & & \\
$\hbar(\E)\to\hbar(\E)\,\theta(\E_h - \E)$, & & & & \\[1mm]
$\mbar(\E)\to\mbar(\E)\,\theta(\E - \E_m)$
                        & 0.2130 & 24.5 \%   & -0.99     & 173.83\\[1mm]\hline
\Eqs{e:CRN0}  with      & & & & \\[1mm]
$M(\E)=\mbar(\E)$, $H(\E)=\hbar(\E)$
                        & 0.2095 & 25.8 \%   & -1.06    & 183.82  \\[1mm]\hline
\Eqs{e:CRN0}  with      & & & & \\[1mm]
$M(\E)=1$, $H(\E)=1$, i.e.
\eqs{e:0020}            & 0.2372    &   16.0 \%      & 2.89   & 193.66\\[1mm]\hline
\Eqs{e:0020a} 
                        & 0.4422      & 57.3  \%      & 18.26  &
                        643.97       \\\hline
\end{tabular}}
}{
 A comparison of the wave speed $C$ and post-front voltage amplitudes
 $\Eomega$ and the maximum rate of AP rise $(\de V / \de t)_\mathrm{max}$ 
 of various approximations to the \citet{CRN98} model. Prior
 to firing, the tissue in the models was set at rest at the standard
 values of the parameters, see \cite{CRN98}.
 In these and other numerical results  $\Curv=0$ is assumed
 unless explicitly stated otherwise. Space-clamped versions of the
 models are used to compute $(\de V / \de t)_\mathrm{max}$.
}{t:01}

\renewcommand{\myfigure}[3]{
\clearpage
\begin{figure}
\begin{center}
#1
\end{center}
\caption{#2}
\label{#3}
\end{figure}
}

\myfigure{
\includegraphics{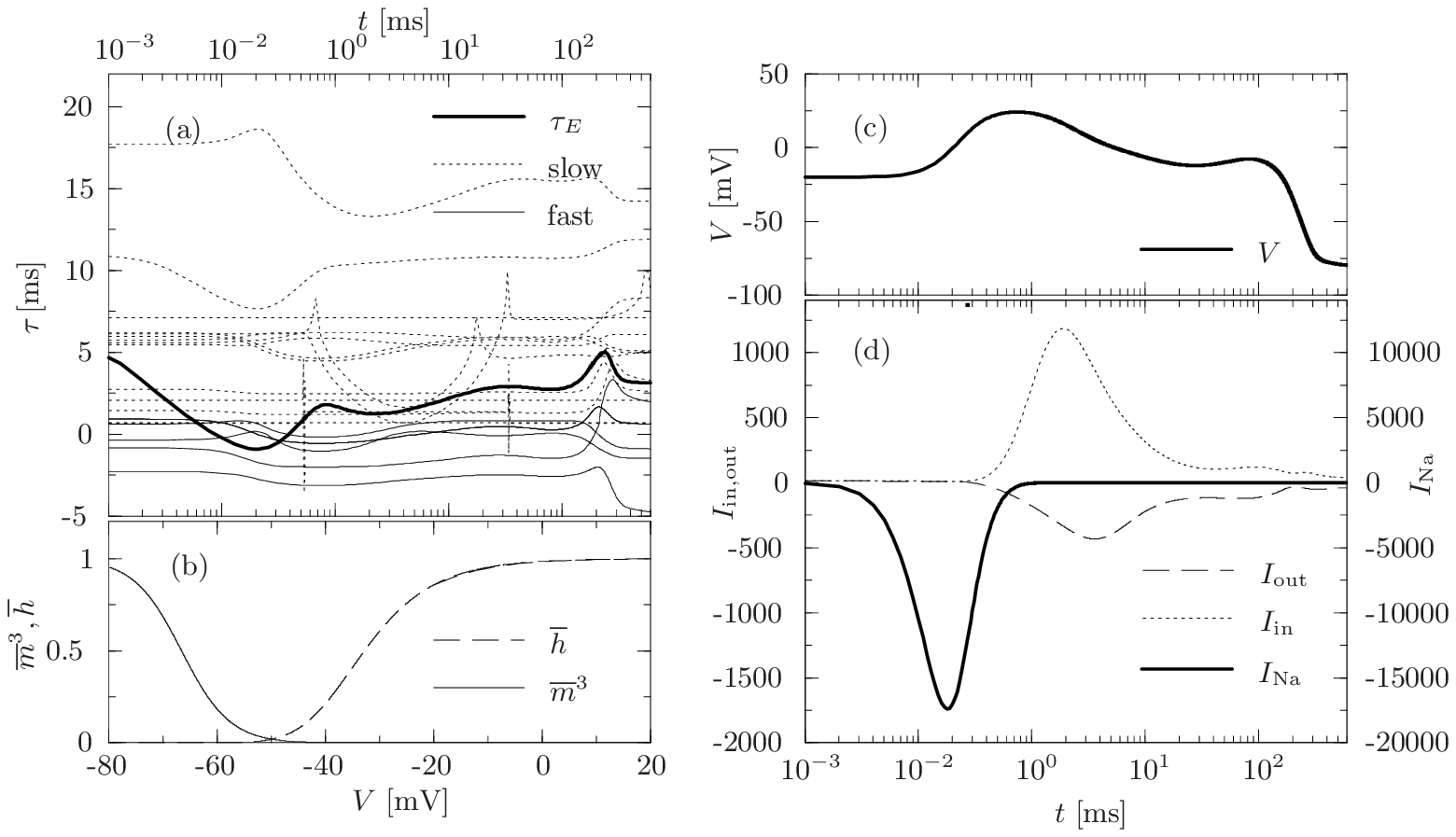}
}{Asymptotic properties of the atrial model of Courtemanche
  \etal{} \cite{CRN98}. (a) Time scale functions of dynamical
  variables vs.~time. (b) Quasistationary values of the gating
  variables $\mbar$ and $\hbar$. (c) Transmembrane voltage $\E$ as a
  function of time. (d) Main  ionic currents vs.~time. 
  $I_{\mathrm{in}} = 
  I_{\mathrm{b,Na}}+I_{\mathrm{NaK}}+I_{\mathrm{Ca,L}}+I_{\mathrm{b,Ca}}+I_{\mathrm{NaCa}}$
  and   $I_{\mathrm{out}} =
  I_{\mathrm{p,Ca}}+I_{\mathrm{K1}}+I_{\mathrm{to}}+I_{\mathrm{Kur}}+I_{\mathrm{Kr}}+I_{\mathrm{Ks}}+I_{\mathrm{b,K}}$
  are the sums of all inward and outward currents, respectively and 
  the individual currents are described in \cite{CRN98}.  
  The results are obtained 
  for a space-clamped version of the model at values of the parameters
  as given in \cite{CRN98}.  In (c) and (d) a
  typical AP is triggered by initialising the
  transmembrane voltage to a non-equilibrium value of $V=-20$~mV.
}{f:0000}

\myfigure{
\includegraphics{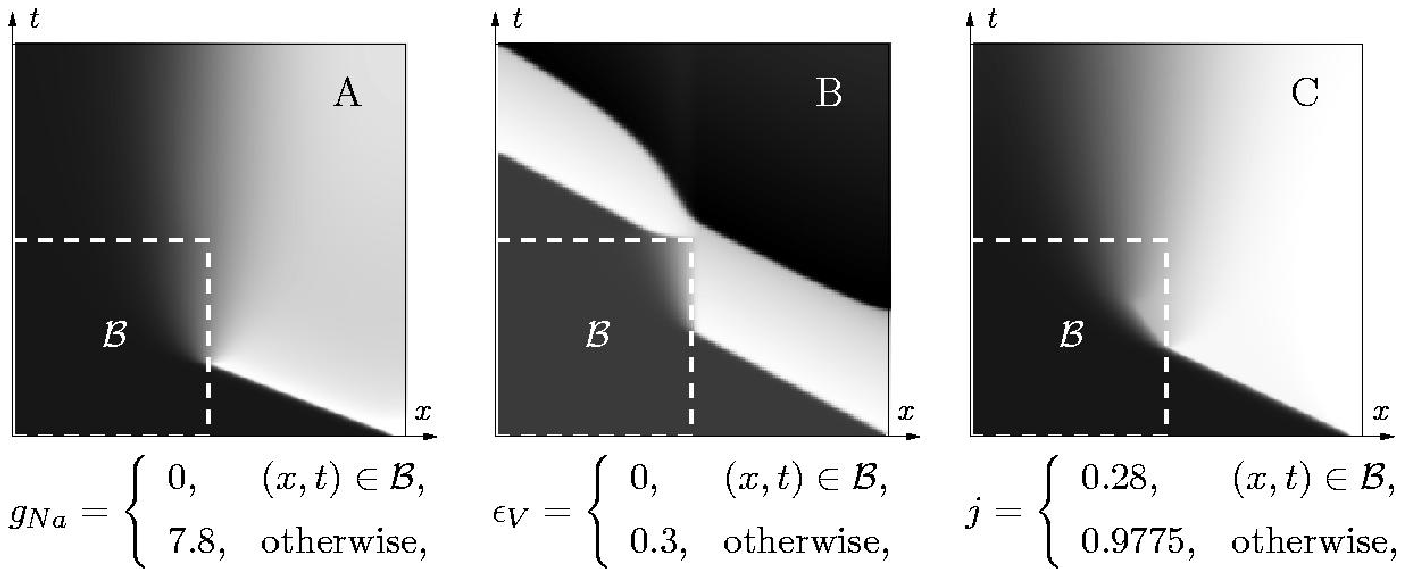}
}{
  Response to a temporary local block of
  excitability ($\Block$) in the models of (A) \citet{CRN98},
  (B) FitzHugh-Nagumo \eqs{FHN} 
  and (C) in \eqs{e:0020}.
  The border of  the blocked region is shown by broken
  lines. Solutions are represented by shades of gray: black is the
  smallest and white is the largest value of $\E$ within the
  solution. The parameters of the FitzHugh-Nagumo model are
  $\beta=0.75$, $\gamma=0.5$ and $\epsilon_\g=0.03$,  while for the
  two other models the same parameter values as described in
  \cite{CRN98} are used, the block is described in the plots.
  The value of $j=0.28$  in the block in (c) is just below the
  propagation threshold, see \fig{f:0060}.  
  The time and space ranges (in dimensionless units) are $70
  \times 70$ in (B)   and $80\times 50$ in (A) and (C).
}{f:0010}

\myfigure{
\includegraphics{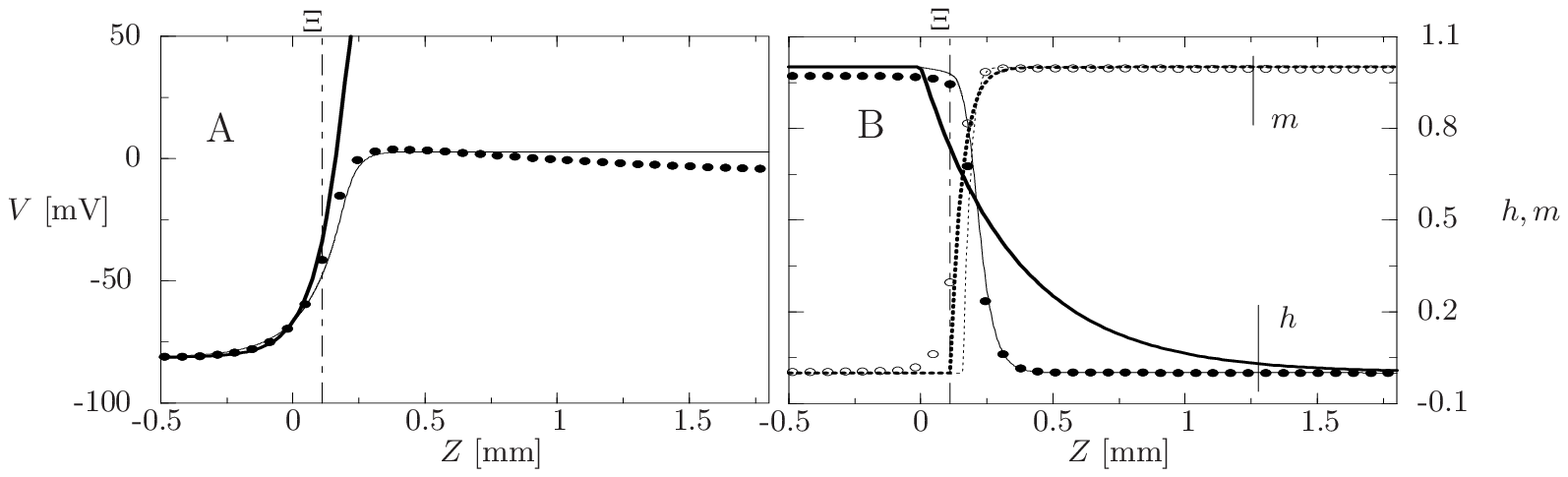}
}{
  (A) The AP potential and (B) the gating variables $h$ and $m$ as
  functions of the travelling wave coordinate $Z=z\sqrt{D}$. The
  solution of the model of \citet{CRN98} is given by circles, of the
  full three-variable model of \eqs{e:0020} by thin lines, and the
  analytical solution given by \eqs{e:0050} for ${\INabar} =
  \INabar(\Em)=781.8$, ${\tau_h}= \tau_h(\Em)=1.077$, ${\tau_m}=
  \tau_m(\Em)=0.131$, $\Ealpha=-81.18$ mV and $j=0.956$ by thick
  lines. The gates $h$ and $m$ are indicated in the plot. The
  position of the internal boundary point $\Xi=\xi\sqrt{D}$ is indicated by a
  dash-dotted line.
}{f:0020}

\myfigure{
\includegraphics{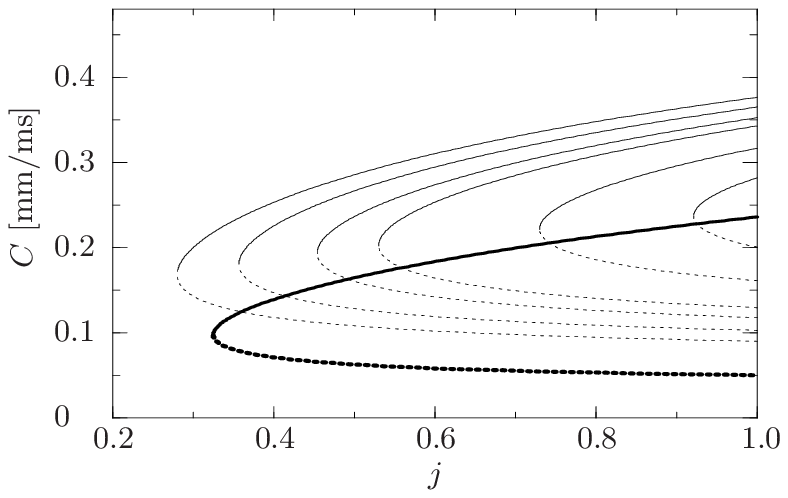}
}{The wave speed $C$ as a function of the excitation
  parameter $j$. Thick lines: the numerical solution of
  \eqs{e:0030}. Thin lines: solution \eq{e:0070} for values
  of ${\tau_h}$ and ${\tau_m}$ corresponding to a 
  selected voltage $\E=\E_*$ in \eqs{e:0030a}. 
  From right to left:
  $\E_*=-28$, $-30$, $\Em$, $-34$, $-36$ $-38$~(mV). 
  In  both cases $\Ealpha=-81.18$~mV and  $\Curv=0$~mm$^{-1}$.
}{f:0030}

\myfigure{
\includegraphics{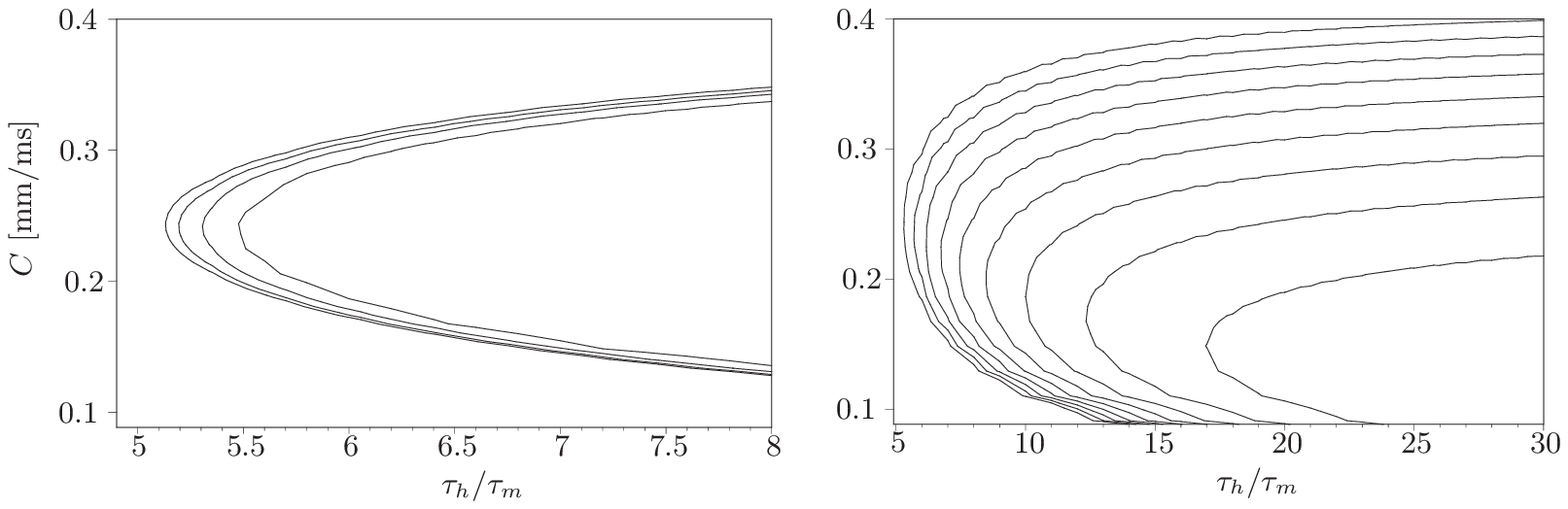}
}{The wave speed $C$ as a function of the time-scale ratio
$\tau_h/\tau_m$ in the caricature model~\eqtwo(e:0030,e:0040). 
The values of $\tau_h$
and $\INabar$ are fixed to the values of the corresponding functions in 
\eqs{e:0030a} at a selected voltage $\E=\E_*$,
the pre-front voltage is $\Ealpha=-81.18$ mV and curvature is $\Curv=0$~mm$^{-1}$.
Left plot: left to right, $\E_*=-38$, $-36$, $-34$ and $-32.7=\Em$ (mV), 
and $j=0.9775$.  
Right plot: right to left, $j=0.2$ to $1.0$ and $\E_*=\Em$.
}{f:0031}

\myfigure{
\includegraphics{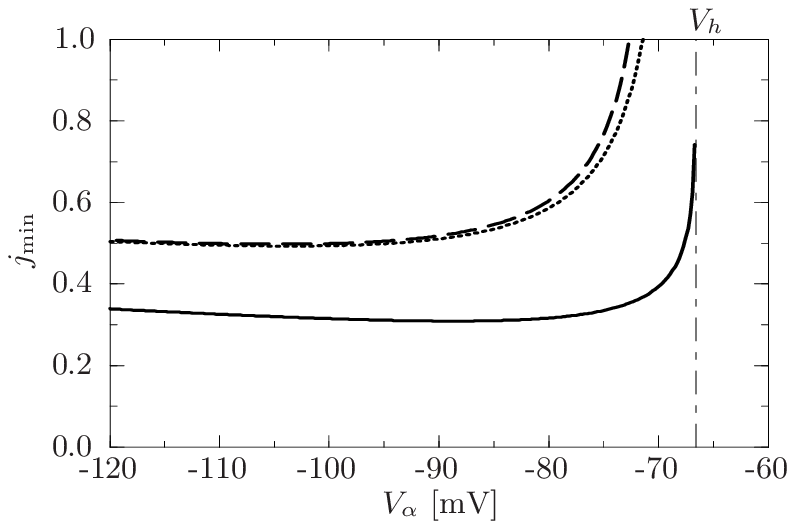}
}{
  The threshold value $\jmin$ above which propagation is
  possible, 
  as a function of the
  pre-front voltage $\Ealpha$ for the same values of the
  parameters as in \fig{f:0020}, \ie{} 
  ${\tau_h}= 1.077$, ${\tau_m}= 0.131$. 
  Shown are different approximations to the perturbation expansion
  given by \eq{e:0090}:
  solid line, zeroth order, \eq{e:0100};
  dashed line, first-order, \eq{e:0101};
  dotted line: second-order.  
}{f:0040}

\myfigure{
\includegraphics{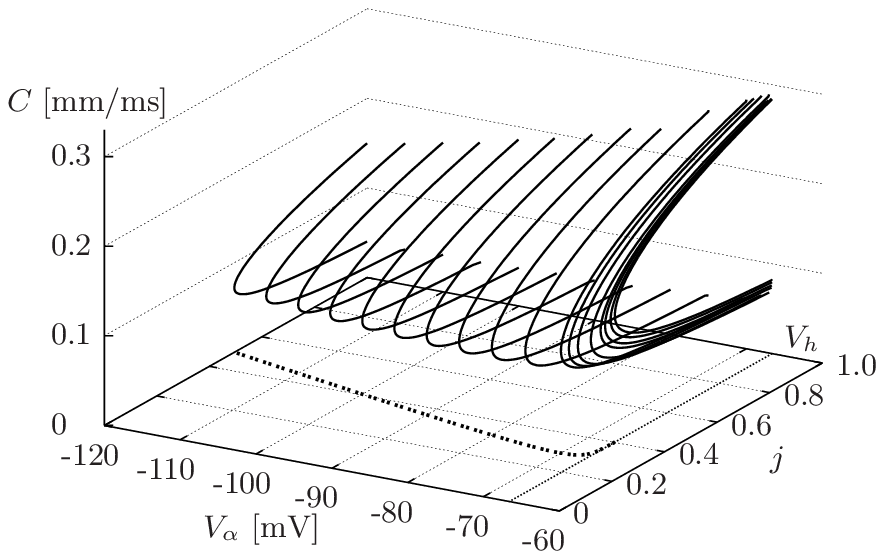}
}{The wave speed $C$ as a function of $j$ and $\Ealpha$,
  for the model of \eqs{e:0030}. Rapid
  changes are indicated by a higher density of curves. The thick
  dotted line on the base represents the threshold value $\jmin$ and may be
  compared to the results in \fig{f:0040}.
}{f:0050}

\myfigure{
\includegraphics{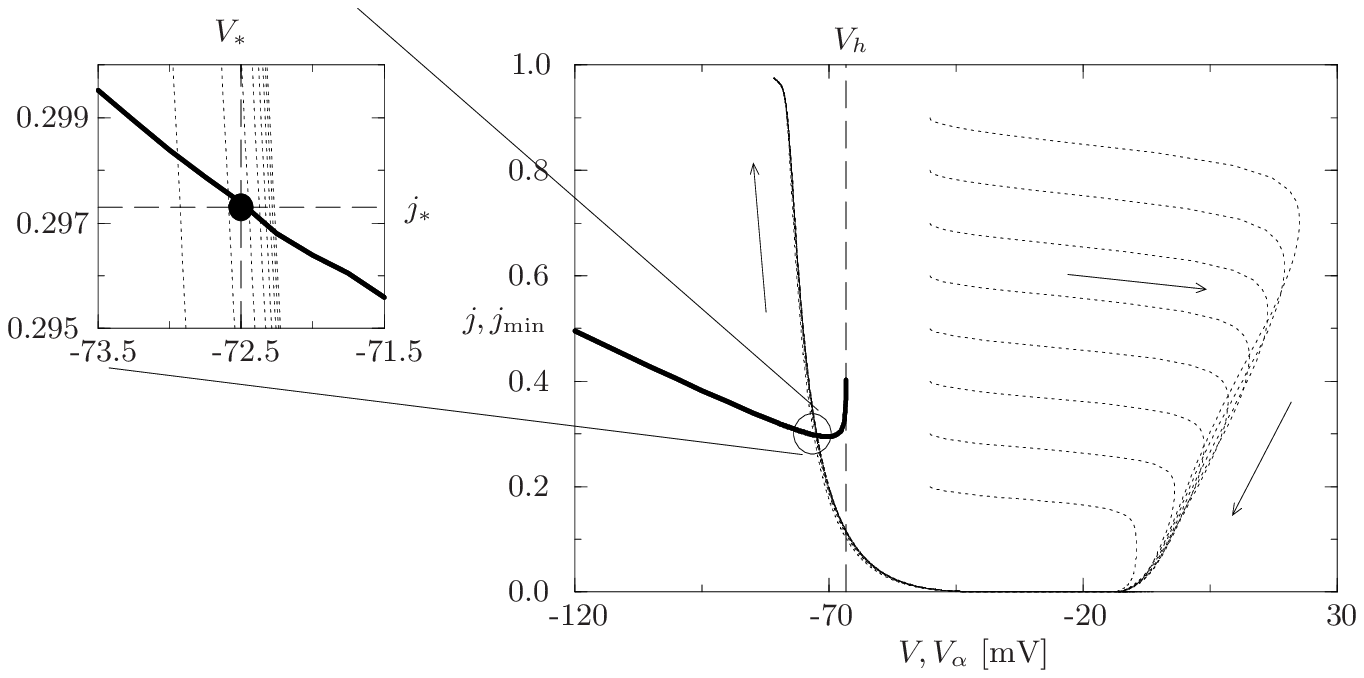}
}{
  The thick solid line represents
  the threshold value $\jmin$ for excitation failure as a function of $\Ealpha$
  for the model given by \eqs{e:0030}.
  The dotted lines represent projections of AP trajectories
  in the space-clamped detailed model of \cite{CRN98}. 
}{f:0060}

\myfigure{
\includegraphics{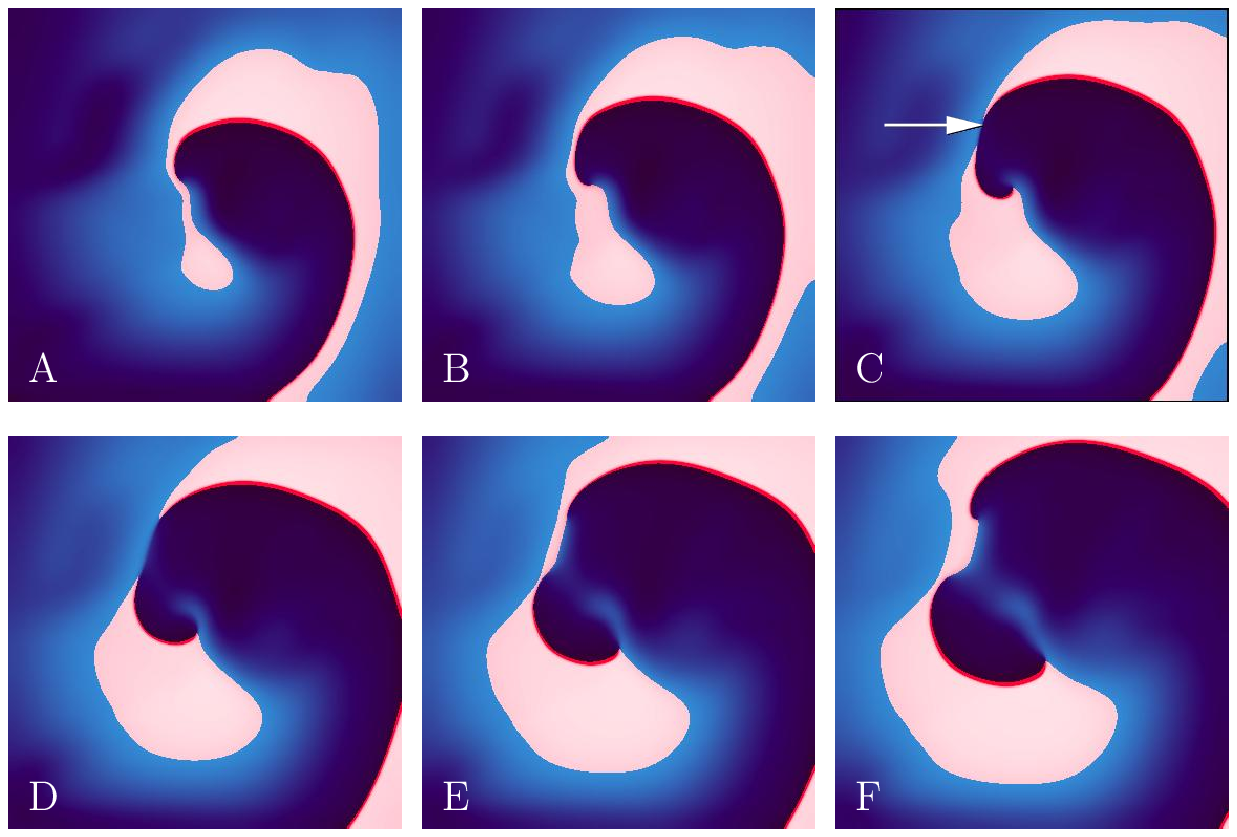}
}{
  Local propagation block, dissipation and break-up of the front of 
  a re-entrant excitation wave.   The density plots represent
  the distribution of the transmembrane voltage $\E$
  (red component) in regions of super-threshold (white) and of
  sub-threshold (blue) excitability $j$. 
  The white arrow indicates the time and place the propagation block begins.
  The time increases from (A) to (F)   with $\Delta t = 20$ ms;
  size of the simulation domain is $75\,\mathrm{mm}\times75\,\mathrm{mm}$.
}{f:0070}

\myfigure{
\includegraphics{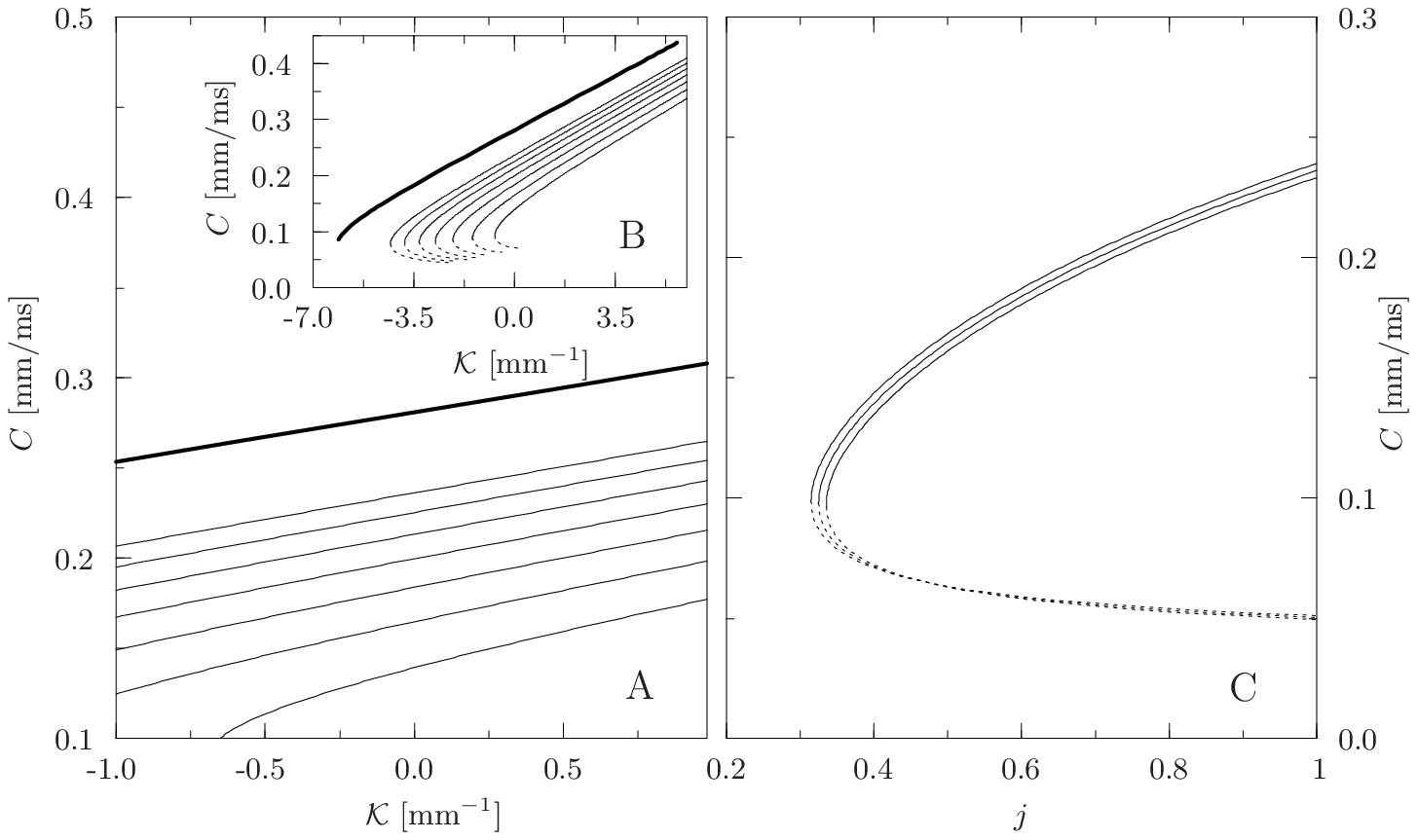}
}{(A) and (B) The wave speed $C$ for the model of
  \eqs{e:0030}, \ref{e:0040} as a function of the curvature for values of $j=1 \dots 
  0.4$ (from top to bottom). Results for the detailed model
  \cite{CRN98} are denoted by thick solid lines. (C)  The wave speed
  $C$ in the model given by 
  \eqs{e:0030} as a function of  $j$ for $\Curv = 0.1$, 0 and
  $-0.1$ mm$^{-1}$ (from top to bottom).
}{f:0080}

\end{document}